\documentclass[onecolumn]{raa}

\usepackage{graphicx,times}             
\usepackage{natbib}
\usepackage{amssymb,amsmath}
\bibpunct{(}{)}{;}{a}{}{,}
\usepackage{booktabs}
\usepackage[T1]{fontenc}
\usepackage{url}
\usepackage{xcolor}
\usepackage{lineno}
\usepackage{tabularx}
\usepackage[colorlinks=true, linkcolor=blue, citecolor=blue, urlcolor=blue]{hyperref}

\begin{document}

  \title{Calibration database design for the wide-field X-ray telescope on board the \textit{Einstein Probe}
}

   \volnopage{Vol.0 (20xx) No.0, 000--000}      
   \setcounter{page}{1}          

   \author{Huaqing Cheng 
      \inst{1}
   \and Hai-Wu Pan
      \inst{1,*}
   \and Yuan Liu
      \inst{1,\dagger}
   \and Donghua Zhao
      \inst{1}
   \and Wenxin Wang
      \inst{1}
   \and He-Yang Liu
      \inst{1}
   \and Chen Zhang
      \inst{1, 2}
   \and Zhixing Ling
      \inst{1, 2, 3}
   \and Weimin Yuan
      \inst{1, 2}
   }

   \institute{National Astronomical Observatories, Chinese Academy of Sciences,
             Beijing 100012, China; {\it panhaiwu@bao.ac.cn, liuyuan@bao.ac.cn}\\
        \and
             School of Astronomy and Space Science, University of Chinese Academy of Sciences, 19A Yuquan Road, Beijing 100049, China\\
        \and
             Institute for Frontier in Astronomy and Astrophysics, Beijing Normal University, Beijing 102206, China\\
\vs\no
   {\small Received 20xx month day; accepted 20xx month day}}

\abstract{The wide-field X-ray telescope (WXT) is the primary payload of the Einstein Probe (EP) mission. Utilizing innovative lobster-eye micro-pore optics (MPO), it achieves an unprecedented combination of a large instantaneous field-of-view (FoV) and high detection sensitivity. Precise scientific data analysis relies fundamentally on a standardized and robust calibration database (CALDB). In this paper, we present the design and implementation of the WXT CALDB.
The database is designed to manage calibration data for 12 independent WXT modules and 48 CMOS sensors, incorporating both extensive on-ground calibration results and dynamic in-orbit updates. 
We detail the directory hierarchy, naming conventions, functionality, and the specific formats of key calibration products. The WXT CALDB is seamlessly integrated into the WXT data analysis pipeline, ensuring high data integrity and maximizing the scientific return of the mission.
}

   \authorrunning{H.~Q.~Cheng et al.}            
   \titlerunning{Calibration database design for the EP-WXT}  

   \maketitle

\section{Introduction}
\label{sect:intro}

The Einstein Probe \citep[EP\footnote{\url{https://ep.bao.ac.cn}},][]{Yuan2015, Yuan2018, Yuan2022, Yuan2025} is an interdisciplinary mission of time-domain and X-ray astronomy, led by the Chinese Academy of Sciences (CAS) in collaboration with the European Space Agency (ESA), the Max-Planck Institute for Extraterrestrial Physics (MPE), and the French Space Agency (CNES). EP aims to perform a thorough and systematic exploration of the soft X-ray sky, utilizing a combination of detection sensitivity and cadence that was inaccessible to previous and current wide-field monitoring missions. Its primary scientific objectives include discovering and characterizing cosmic X-ray transients, such as high-redshift gamma-ray bursts (GRBs), supernova shock breakouts (SBOs), tidal disruption events (TDEs), as well as the electromagnetic counterparts of gravitational wave (GW) events.
Leveraging its large instantaneous field of view (FoV) and high sensitivity, EP will also perform long-term X-ray monitoring of known sources such as X-ray binaries (XRBs) and active galactic nuclei (AGNs). Since its successful launch in January 2024, a wealth of scientific discoveries have been made across various fields of high-energy astrophysics \citep[e.g.,][]{Yin2024,EP240315a, EP240414a, Marino2025, EP240904a, 2025Jin, 2025Li0702}. 
A summary of the metrics and statistics for papers on the Astrophysics arXiv\footnote{\url{https://arxiv.org/archive/astro-ph}} over the year 2025 found that, despite not being among the most used instruments, EP publications are highly cited, with three of the four citation indices being the highest among all telescopes and facilities \citep{Lewis2026arxiv}.

On board the EP satellite are two primary scientific payloads: the Wide-field X-ray Telescope \citep[WXT,][]{2025ChengWXTcalib} and the Follow-up X-ray Telescope \citep[FXT,][]{Chen2020FXT}. The WXT is an advanced instrument featuring novel lobster-eye micro-pore optics \citep[MPO, e.g.,][]{Angel1979,Fraser1992,Fraser1993,Willingale1998,2017ZhaoDH_simulation,Willingale2016, 2022ZhangChenApJL,2023LingZXRAA, Cheng2024}, which combines a large instantaneous FoV of over 3600 square degrees with a high detection sensitivity of $\sim(2-3)\times10^{-11}\mathrm{~erg~s^{-1}~cm^{-2}}$ within a single snapshot of $1000\mathrm{~s}$ in exposure. It consists of 12 identical modules, each containing 36 MPO plates and four large-format, back-illuminated scientific complementary metal-oxide semiconductor \citep[CMOS, e.g.,][]{2022WangWXa,2022WangWXb,Wu2022,Wu2023pasp1,Wu2023pasp2,Wu2023nima,LiuMJ2023, ChenMX2024, LiuMJ2025} sensors as the focal plane detectors.
To the best of our knowledge, EP-WXT is the first implementation of mass-produced MPO technology and CMOS detectors in a dedicated X-ray astronomy mission.
The specifications of the WXT payload are detailed in Table \ref{tab:wxt_specifications}.

\begin{table*}[!htbp]
\caption{Specifications of the WXT Instruments.}
\centering
\begin{tabular}{ll}
\toprule
\textbf{Parameters} & \textbf{Values} \\
\midrule
Number of modules & 12 \\
Optic design & Lobster-eye MPO \\
Detector & CMOS \\
Field of view (FoV) & $>3600$ square degrees \\
Focal length & $375\mathrm{~mm}$ \\
Source positioning accuracy & $\lesssim 2^{\prime}$ (J2000, 90\% C.L.) \\
Effective area & $2$--$3\mathrm{~cm^2}$ @ $1.25\mathrm{~keV}$ \\
Spatial resolution & $3^{\prime}$--$5^{\prime}$ @ $1\mathrm{~keV}$ \\
Effective energy band & $0.5$--$4\mathrm{~keV}$ \\[1ex]

Energy resolution & $120$--$140\mathrm{~eV}$ @ $1.25\mathrm{~keV}$ \\
 & $150$--$170\mathrm{~eV}$ @ $3\mathrm{~keV}$ \\
 & $180$--$200\mathrm{~eV}$ @ $4.5\mathrm{~keV}$ \\[1ex] 

Limiting flux ($\mathrm{erg~s^{-1}~cm^{-2}}$) & $\sim8.9\times10^{-10}$ ($27.6\mathrm{~mCrab}$) @ $10\mathrm{~s}$ \\
 & $\sim1.2\times10^{-10}$ ($3.9\mathrm{~mCrab}$) @ $100\mathrm{~s}$ \\
 & $\sim2.6\times10^{-11}$ ($0.8\mathrm{~mCrab}$) @ $1000\mathrm{~s}$ \\[1ex]

Readout noise & $\sim 3~\mathrm{e}^{-}$ \\
Time resolution & $50\mathrm{~ms}$ \\
\bottomrule
\end{tabular}
\label{tab:wxt_specifications}
\end{table*}

In X-ray astronomy, the derivation of accurate physical properties of celestial sources from raw telemetry data relies fundamentally on a standardized calibration database \citep[CALDB, see][for a detailed introduction]{Arnaud2011handbook}. 
The primary objective of the WXT CALDB is to systematically organize and archive the calibration results obtained from on-ground experiments and continuous in-orbit monitoring. During the design and development phase, several core requirements were prioritized. First, the database architecture must seamlessly integrate with the WXT data reduction pipeline (Liu et al. in preparation) to ensure the generation of high-fidelity scientific products. Second, to facilitate data exploitation by international astronomers, the WXT CALDB is strictly designed to comply with the format conventions established by the High Energy Astrophysics Science Archive Research Center (HEASARC)\footnote{\url{https://heasarc.gsfc.nasa.gov/docs/heasarc/caldb/}} and the Office of Guest Investigator Programs (OGIP)\footnote{\url{https://hesperia.gsfc.nasa.gov/rhessidatacenter/software/ogip/ogip.html}}. This standardization ensures that the broader astronomical community can analyze WXT observations utilizing widely adopted tools and methodologies.

While the WXT CALDB strictly adheres to these universal standards, its practical implementation still faces distinctive challenges arising from WXT's unique instrumental characteristics.
First, the WXT CALDB must systematically parameterize the complex optical responses of the lobster-eye MPO and manage a massive detector array. As the first large-scale implementation of MPO technology in space, WXT produces a characteristic cruciform point spread function (PSF) and distinct vignetting effects \citep{Fraser1992, 2017ZhaoDH_simulation}. The CALDB is demanded to model these responses across varying energies and positions, simultaneously managing a large volume of files for 48 CMOS sensors across 12 modules.
Second, the CALDB must incorporate a sophisticated spatial correction framework to achieve the designed $\leq2'$ source localization. Realizing this precision across the entire FoV of over 3600 square degrees is practically unprecedented, which requires a reliable management of complex geometric transformations among the 12 modules and modeling of inherent non-linear spatial distortions from the MPO manufacturing process \citep{2022ZhangChenApJL, 2026ChengLEIA}, posing a novel architectural challenge for spatial calibration storage.
Third, the WXT CALDB requires a highly adaptable, time-dependent architecture to track the in-orbit evolution of the overall instrumental performance. While tracking optical degradation (e.g., effective area deterioration due to accumulative contamination) is routine for space missions \citep[e.g.,][]{Marshall2004, Grant2024}, WXT marks the pioneering massive application of large-format scientific CMOS detectors in space-borne X-ray astronomy. To address their uncharted long-term behavior under space radiation and thermal fluctuations, the CALDB utilizes multi-dimensional grids interpolating across observation time and temperature. This dynamic framework accurately traces continuous detector evolution (e.g., gain, spectral resolution and defects), providing a crucial benchmark for future CMOS missions.

In this paper, we describe the design and implementation of the WXT CALDB. Section 2 describes the overall architecture and naming conventions complying with the HEASARC standards; Section 3 details the specific file formats and extraction mechanisms for key calibration products; and Section 4 summarizes the current status and future updates of the database.

\section{WXT Calibration Database Architecture} 
\label{sect:arch}
\label{sec:wxtcaldb_arch}

\subsection{Directory Structure}

The WXT CALDB is organized into a hierarchical directory structure that ensures logical separation between basic calibration parameters and high-level response products. Specifically, the root directory, named \texttt{CALDB/data/ep/wxt}, contains the master index file \texttt{caldb.indx}, the sub-directory \texttt{index} (storing calibration index file of different epochs), and two primary sub-directories: \texttt{bcf} (storing Basic Calibration Files) and \texttt{cpf} (storing Calibration Product Files). The \texttt{bcf} directory comprises eight specialized sub-folders, including \texttt{teldef}, \texttt{badpix}, \texttt{bias}, \texttt{effarea}, \texttt{ftrans}, \texttt{gain}, \texttt{grade}, and \texttt{qe}. The \texttt{cpf} directory includes five sub-folders for analysis-ready products: \texttt{arf}, \texttt{bkg}, \texttt{psf}, \texttt{vign}, and \texttt{rmf}. The datatypes and short descriptions of the WXT calibration file set (WCFS) are detailed in Table \ref{tab:wcfs_files}. 

\begin{table*}[!htbp]
\caption{Datatypes and Short Description of WCFS Files}
\centering
\begin{tabular}{llll}
\toprule
\textbf{Root Directory} & \textbf{Level 1} & \textbf{Level 2} & \textbf{Description} \\
\midrule
\texttt{\$CALDB/data/ep/wxt} & \texttt{bcf} & \texttt{teldef} & Telescope definition file \\
 &  & \texttt{badpix} & Table of dead/hot pixels applied to data and masked \\
 &  & & on board \\
 &  & \texttt{bias} & Detector bias \\
 &  & \texttt{effarea} & Mirror effective area \\
 &  & \texttt{ftrans} & Filter transmission curves \\
 &  & \texttt{gain} & Conversion factors from digitized signals to Pulse \\
 &  & & Invariant (PI) channels \\
 &  & \texttt{grade} & Grade definition \\
 &  & \texttt{qe} & Quantum efficiency \\
\cmidrule{2-4}
 & \texttt{cpf} & \texttt{arf} & On-axis Ancillary Response File for standard extraction\\
 &  &  & region \\
 &  & \texttt{bkg} & Background spectra \\
 &  & \texttt{psf} & Point Spread Function correction factor \\
 &  & \texttt{rmf} & Response matrices for PI spectra \\
 &  & \texttt{vign} & Vignetting correction factor \\
\midrule
 & \texttt{index} & \texttt{caldb.indxYYY} & Calibration index file of different epochs \\
 &  & \texttt{YMMDD} & \\
\midrule
 & \texttt{caldb.indx} & & A link to the current calibration index file in the \\
 &  & & \texttt{index/} subdirectory \\
\bottomrule
\end{tabular}
\label{tab:wcfs_files}
\end{table*}

\subsection{File Naming Conventions}
To accommodate the large-scale sensor array of WXT, a standardized naming convention is implemented: \texttt{epw<cmos\_number><datatype>\_YYYYMMDDv<version>.<ext>}. Here, \texttt{<cmos\_number>} ranges from 1 to 48, uniquely identifying each of the 48 CMOS detectors across the 12 WXT modules. The \texttt{<datatype>} identifies the calibration content (e.g., \texttt{badpix}, \texttt{gain}), while \texttt{YYYYMMDD} and \texttt{v<version>} facilitate temporal indexing and version control. \texttt{<ext>} denotes the type of the calibration file.

\subsection{Indexing and Metadata}
The indexing mechanism relies on mandatory FITS keywords embedded in each file's primary header. Key parameters include \texttt{TELESCOP} (`EP'), \texttt{INSTRUME} (`WXT'), and \texttt{DETNAM} (identifying the specific CMOS sensor from 1 to 48). Temporal validity is governed by the \texttt{CVSD} (Start Date) and \texttt{CVST} (Start Time) keywords, allowing the data reduction pipeline to automatically select the calibration file most appropriate for a given observation epoch.

\begin{table}[!htbp]
\caption{WCFS Mandatory Header Keywords}
\centering
\begin{tabular}{lll}
\toprule
\textbf{Keyword Name} & \textbf{Keyword Value} & \textbf{Comment} \\
\midrule
\texttt{TELESCOP} & `EP' & Telescope (mission) name \\
\midrule
\texttt{INSTRUME} & `WXT' & Instrument name \\
\midrule
\texttt{DETNAM}   & `CMOS<cmos\_number>' & CMOS number <1 to 48> \\
\midrule
\texttt{DATE}     & YYYY-MM-DDThh:mm:ss & Creation date \\
         &                       & This keyword is omitted for empty primary headers. \\
\midrule
\texttt{CHECKSUM} & up to date checksum & HDU checksum updated date \\
\midrule
\texttt{DATASUM}  & up to date data unit checksum & Data unit checksum updated date \\
\bottomrule
\end{tabular}
\label{tab:header_keywords}
\end{table}

\begin{table}[!htbp]
\caption{WCFS Table Headers Mandatory Keywords}
\centering
\begin{tabular}{lll}
\toprule
\textbf{Keyword Name} & \textbf{Keyword Value} & \textbf{Comment} \\
\midrule
\texttt{EXTNAME} & <extension name> & Name of the binary table extension or \\
        &                & name of the image extension. Omitted if \\
        &                & data are stored in the primary header. \\
\midrule
\texttt{ORIGIN}  & <organization name> & Source of FITS file \\
\midrule
\texttt{CREATOR} & <task name and version number> & Creator \\
\midrule
\texttt{CONTENT} & <short description of the content> & File content \\
\midrule
\texttt{FILENAME} & <file name> & File name \\
\midrule
\texttt{VERSION} & <version number> & Extension version number \\
\midrule
\multicolumn{3}{c}{\textbf{CALDB Keywords:}} \\
\midrule
\texttt{CCLSxxxx} & OGIP-class of calibration file & Dataset is a Calibration Product File / \\
         &                                & Dataset is a Basic Calibration File \\
\midrule
\texttt{CDTPxxxx} & <datatype code> & Calibration file contains data \\
\midrule
\texttt{CCNMxxxx} & <extension codename> & Type of calibration data \\
\midrule
\texttt{CDESxxxx} & <descriptive string> & Description \\
\midrule
\texttt{CVSDxxxx} & <start valid data> & UTC date when file should first be used \\
\midrule
\texttt{CVSTxxxx} & <start valid time> & UTC time when file should first be used \\
\bottomrule
\end{tabular}
\label{tab:caldb_keywords}
\end{table}

\begin{table}[!htbp]
\caption{WCFS Table Headers Keywords Required Under Certain Circumstances}
\centering
\begin{tabular}{lll}
\toprule
\textbf{Keyword Name} & \textbf{Keyword Value} & \textbf{Comment} \\
\midrule
\texttt{CBDnxxxx} & array describing parameter limitations & / Parameter boundary \\
         & of the dataset & \\
\midrule
\texttt{CSYSNAME} & spatial coordinate system in use & / spatial coord system used in this dataset \\
\midrule
\texttt{TDIMnnn}  & Number of elements \& Ordering of $n$-d array & / Array dimensions \\
\midrule
\texttt{HDUCLASS} & `OGIP' & / format conforms to OGIP standards \\
\midrule
\texttt{HDUDOC}   & <document number> & / Document describing the format \\
\midrule
\texttt{HDUCLASn} & <character string to classify the extension> & / (Specific to the type) \\
\midrule
\texttt{HDUVERSn} & <string giving the format version> & / Version of file format \\
\midrule
\texttt{TIMESYS}  & TT & / Time system \\
\midrule
\texttt{MJDREFI}  & 58849 (TBD) & / Reference MJD, integer part \\
\midrule
\texttt{MJDREFF}  & 7.4287037e-4 & / Reference MJD, fractional part \\
\midrule
\texttt{CLOCKAPP} & F & / If clock corrections are applied (F/T) \\
\bottomrule
\end{tabular}
\label{tab:conditional_keywords}
\end{table}

\section{Calibration Data Products and Formats} 
\label{sect:products}

\subsection{Basic Calibration Files}
\label{sec:bcf}
The Basic Calibration Files (BCF) encapsulate the fundamental physical and electronic characteristics of the telescope and detectors. These files are typically generated from on-ground calibrations and refined by in-orbit monitoring.

\subsubsection{Telescope Definition (\texttt{teldef})}
\label{sec:teldef}

The telescope definition file (\texttt{teldef}) governs the spatial mapping sequence, translating telemetry coordinates (\texttt{RAWX}, \texttt{RAWY}) into the physical detector frame (\texttt{DETX}, \texttt{DETY}), and ultimately projecting them onto the celestial sphere (\texttt{SKY}). The file is named as \texttt{epw<cmos\_number>\_YYYYMMDDvNNN.teldef}, containing one primary header data unit (HDU) and a table extension named \textbf{\texttt{NONLINEAR}}.
The primary HDU hosts the fundamental metadata specifying the coordinate hierarchies, instrument characteristics, and their respective alignment matrices in its header keywords.

For the EP-WXT pipeline, three distinct spatial reference systems are established. Consequently, the \texttt{NCOORDS} parameter is fixed at 3, with the system hierarchy defined as follows:
\begin{itemize}
    \item \texttt{COORD0 = `RAW'}
    \item \texttt{COORD1 = `DET'}
    \item \texttt{COORD2 = `SKY'}
\end{itemize}

The native \texttt{RAW} coordinates, extracted directly from telemetry, encompass the $4096 \times 4096$ pixel array of the CMOS sensor. The intermediate \texttt{DET} coordinate plane maintains this exact dimensional footprint. Conversely, the \texttt{SKY} coordinate grid is expanded to a $10000 \times 10000$ array. This extended canvas is necessary to fully contain the projected detector footprint, particularly when accommodating a 45-degree roll angle relative to the Right Ascension and Declination (RA/Dec) axes.

The mathematical translation from the \texttt{RAW} to the \texttt{DET} frame proceeds through a two-phase linear transformation:

1) Initially, telemetry coordinates are mapped onto an intermediate internal plane. While this step was historically designed to support focal planes with multiple sub-units, it is preserved here for standard pipeline compatibility. The intermediate coordinates ($X_{\mathrm{int}}, Y_{\mathrm{int}}$) are calculated as:
\begin{align}
    X_{\mathrm{int}} &= \mathrm{COE\_X\_B} \cdot \mathrm{RAWX} + \mathrm{COE\_X\_C} \cdot \mathrm{RAWY} + \mathrm{COE\_X\_A} \\
    Y_{\mathrm{int}} &= \mathrm{COE\_Y\_B} \cdot \mathrm{RAWX} + \mathrm{COE\_Y\_C} \cdot \mathrm{RAWY} + \mathrm{COE\_Y\_A}
\end{align}
Given that the WXT focal plane utilizes a monolithic CMOS sensor, the transformation coefficients in the \texttt{teldef} header are configured to represent a direct 1:1 mapping:
\begin{itemize}
    \item \texttt{COE\_X\_A = 0}, \texttt{COE\_X\_B = 1}, \texttt{COE\_X\_C = 0}
    \item \texttt{COE\_Y\_A = 0}, \texttt{COE\_Y\_B = 0}, \texttt{COE\_Y\_C = 1}
\end{itemize}
which simplifies the equations to, 
\begin{align}
    X_{\mathrm{int}} &= \mathrm{RAWX}\\
    Y_{\mathrm{int}} &= \mathrm{RAWY}
\end{align}
2) Subsequently, these internal coordinates are shifted and scaled into the \texttt{DET} coordinate space, using the following relations:
\begin{align}
    \mathrm{DETX} &= \mathrm{DET\_XCEN} + \mathrm{DETXFLIP} \cdot \frac{X_{\mathrm{int}} - \mathrm{INT\_XCEN} - \mathrm{DET\_XOFF}}{\mathrm{DET\_SCAL}} \\
    \mathrm{DETY} &= \mathrm{DET\_YCEN} + \mathrm{DETYFLIP} \cdot \frac{Y_{\mathrm{int}} - \mathrm{INT\_YCEN} - \mathrm{DET\_YOFF}}{\mathrm{DET\_SCAL}}
\end{align}
\begin{align}
    \mathrm{DET\_XCEN} &= \mathrm{DETXPIX1} + \frac{\mathrm{DET\_XSIZ} - 1}{2.0} \\
    \mathrm{DET\_YCEN} &= \mathrm{DETYPIX1} + \frac{\mathrm{DET\_YSIZ} - 1}{2.0}
\end{align}
where the baseline constants are predefined as:
\begin{itemize}
    \item \texttt{INT\_XCEN = DET\_XCEN}
    \item \texttt{INT\_YCEN = DET\_YCEN}
    \item \texttt{DET\_XOFF = 0}
    \item \texttt{DET\_YOFF = 0}
    \item \texttt{DETXFLIP = 1}
    \item \texttt{DETYFLIP = 1}
    \item \texttt{DET\_SCAL = 1}
    \item \texttt{DETXPIX1 = 1}
    \item \texttt{DETYPIX1 = 1}
    \item \texttt{DET\_XSIZ = 4096}
    \item \texttt{DET\_YSIZ = 4096}
\end{itemize}
For each CMOS sensor, the parameters \texttt{DET\_XOFF} and \texttt{DET\_YOFF} denote the positions of the curvature center within the detector coordinate frame, and are obtained from on-ground measurements.

Finally, the projection from \texttt{DET} coordinates onto the \texttt{SKY} frame relies on a separate transformation. A $3 \times 3$ alignment matrix embedded within the \texttt{teldef} header mathematically ties the detector's physical orientation to the spacecraft's pointing axes. Auxiliary constants essential for astrometry, including CMOS pixel dimensions and the effective focal length, are also cataloged in this header.

\begin{figure*}[!htbp]
    \centering
    \includegraphics[width=\textwidth]{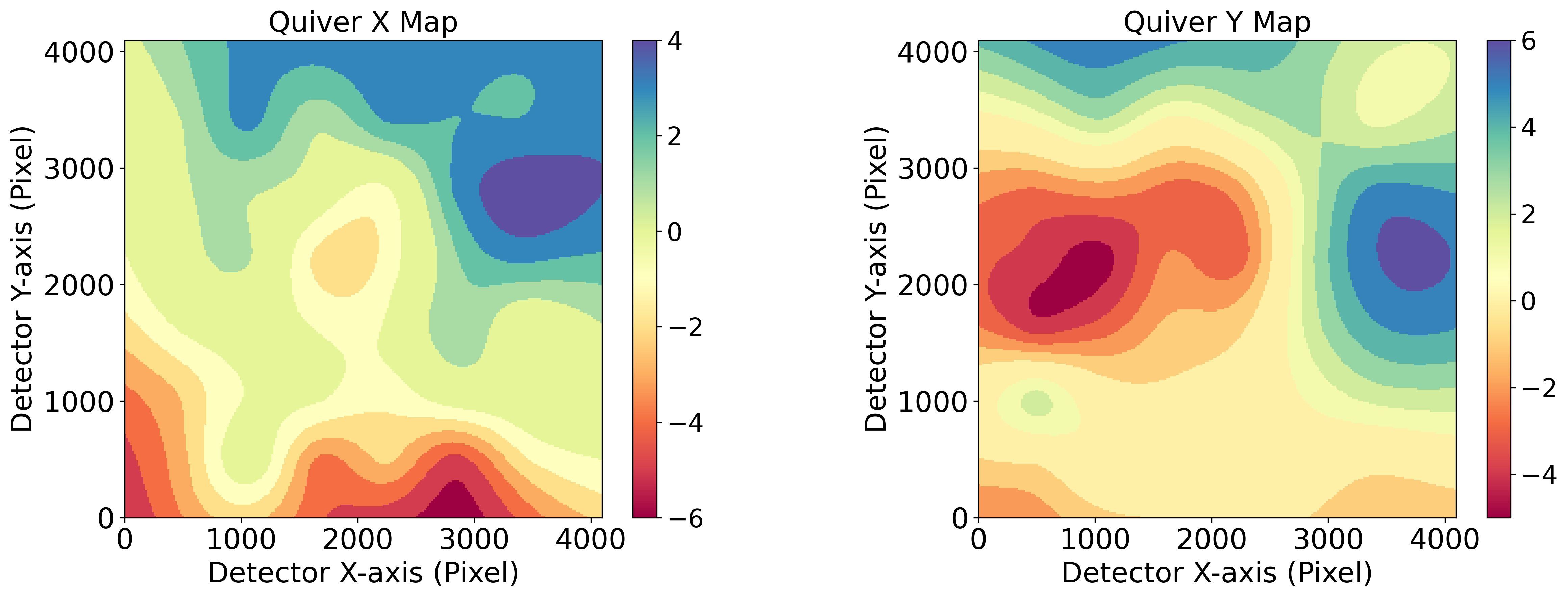}
    \caption{The non-linear correction matrices applied for coordinate transformations. The \textit{Left} panel illustrates the quiver map adjustments for the RAWX coordinate, while the \textit{Right} panel details the corresponding corrections for the RAWY coordinate. The unit is provided in pixel.}
    \label{fig:teldef_nonlinear}
\end{figure*}

A critical, instrument-specific feature of the WXT \texttt{teldef} file is the implementation of a non-linear spatial correction, applied to the original detector coordinates prior to the aforementioned linear transformations. This correction mitigates structural aberrations inherent to the manufacturing of the MPO plates. The corrective data is stored within the first FITS extension (named \textbf{\texttt{NONLINEAR}}) as two distinct $257 \times 257$ quiver matrices, providing independent offset estimations for the \texttt{RAWX} and \texttt{RAWY} coordinates, respectively.
Due to the limitations in performing source localization experiments in the ground stage, these two matrices are initially populated with zeros. 
The matrices are updated during the in-orbit calibration process during the commissioning phase.
Specifically, by performing a series of pointed observations across the entire field of view (FoV), the positions of the focal spot are compared against their theoretical expectations, which are derived from the satellite quaternion and transformation matrix from celestial to detector coordinates. 
The overall positional offsets are minimized by refining the transformation matrix (note that this update will be written into the $3\times3$ alignment matrix mentioned above), leaving a residual offset matrix with an anisotropic pattern \citep[one may refer to Section 4 of][for a detailed illustration]{2026ChengLEIA}. These positional offsets are then smoothed into a $257\times257$ matrix utilizing a radial basis function interpolator (\texttt{scipy.interpolate.RBFInterpolator}). 
An illustrative example of this non-linear correction matrix is presented in Figure \ref{fig:teldef_nonlinear}, with the \textit{Left} and \textit{Right} panels denoting the discrete corrections for the \texttt{RAWX} and \texttt{RAWY} axes, respectively.

\subsubsection{Bad Pixel Calibration File}
\label{sec:badpix}

To manage detector defects, the WXT CALDB maintains two parallel sets of bad pixel files for each of the 48 CMOS sensors. The on-board version (\texttt{epw<cmos\_number>\_onboardbpYYYYMMDDvNNN.fits}) is uploaded to the spacecraft for real-time bad-pixel filtering, while the ground-processing version (\texttt{epw<cmos\_number>\_badpixYYYYMMDD.fits}) is utilized by the WXT on-ground data analysis pipeline. Discrepancies may exist between these versions, as the ground database is continuously updated with newly identified defects prior to their synchronization with the on-board processor.

Both files share an identical FITS structure, featuring a standard null primary header and a binary table extension (named \textbf{\texttt{BADPIX}}) designed to track the spatial and temporal evolution of the CMOS detectors. Each row in the table registers a specific defect through the following key columns:
\begin{itemize}
    \item \texttt{TIME}: The epoch from which the pixel or column is officially flagged as defective. This temporal tag allows the data reduction software to dynamically apply bad-pixel masking appropriate for the specific observation date.
    \item \texttt{RAWX} \& \texttt{RAWY}: The native detector coordinates locating the defect.
    \item \texttt{TYPE} \& \texttt{YEXTENT}: \texttt{TYPE} distinguishes between an isolated pixel defect (value of 1) and a column defect (value of 2). For the latter, \texttt{YEXTENT} specifies the consecutive number of degraded pixels along the Y-axis.
    \item \texttt{BADFLAG}: A 16-bit binary integer indicating the provenance of the defect. For instance, the bitmask sets the second least significant bit (\texttt{b...010}) for defects originated from the on-board table, whereas the least significant bit (\texttt{b...001}) uniquely flags defects discovered during subsequent ground-based analysis.
\end{itemize}

\subsubsection{Bias Calibration Files}
\label{sec:bias}

The bias calibration files record the inherent bias baselines of the CMOS detectors, which must be subtracted during data reduction.
The filename is \texttt{epw<cmos\_number>\_biasYYYYMMDDvNNN.fits}.
Due to onboard storage and processing constraints, the satellite software only subtracts the row-median bias. The residual two-dimensional bias values are subsequently removed by the ground-based analysis software. While the initial bias levels are determined from on-ground calibration, these parameters are updated based on routine in-orbit calibration observations to track temporal variations.

Structurally, the bias FITS file incorporates a standard null primary header followed by two data extensions:
\begin{itemize}
    \item \textbf{\texttt{ROW\_BIAS} extension}: A binary table that stores the row-median bias applied onboard. It contains the \texttt{RAWY} column, indicating the raw spatial row coordinate, and the \texttt{ROW\_BIAS} column, which records the corresponding median subtraction value in Digital Numbers (DN).
    \item \textbf{\texttt{BIAS} extension}: A 2D image array ($4096 \times 4096$) containing the residual bias value for every individual pixel that needs to be further subtracted from the telemetry data.
\end{itemize}

To prevent the truncation of negative noise fluctuations during the onboard median subtraction, an artificial positive offset ($40\mathrm{~DN}$, designated by the \texttt{BIAS0} keyword in the header of the \textbf{\texttt{BIAS} extension}) is systematically added. Therefore, the Pulse Height Amplitude (PHA) of a telemetered photon event is calculated as:
\begin{equation}
    \mathrm{PHA} = \mathrm{raw~PHA} - \mathrm{row~median~bias} + \mathrm{offset}
\end{equation}

Accordingly, the pixel values stored within the \texttt{BIAS} image extension represent the rounded results of the following calculation:
\begin{equation}
    \mathrm{BIAS} = \mathrm{raw~bias~(averaged~from~several~frames)} - \mathrm{row~median~bias} + \mathrm{offset}
\end{equation}

\subsubsection{Gain Calibration File}
\label{sec:gain}

The conversion of telemetered PHA signal into physically standardized Pulse Invariant (PI) energies is realized by digesting the information maintained in the detector gain calibration files. These files encapsulate the intrinsic Energy-Channel (E-C) relation, detailing how the CMOS sensor's charge collection efficiency responds to incident photon energies. The naming convention is \texttt{epw<cmos\_number>\_gainYYYYMMDDvNNN.fits}. The file consists of a standard null primary header and a binary table extension (named \textbf{\texttt{Gain}}) containing the E-C relation data.

To accommodate potential spatial non-uniformities of the E-C relation across the detector plane, the generalized PI transformation is mathematically modeled as a position-dependent linear function:
\begin{equation}
    \mathrm{PI} = \frac{ \mathrm{PHA} \cdot (G_{C0} + x G_{C1} + y G_{C2}) + (G_{C3} + x G_{C4} + y G_{C5}) }{ G_{\mathrm{nom}} }
\end{equation}
where $x$ and $y$ denote the two raw detector coordinates (\texttt{RAWX}, \texttt{RAWY}) of the photon event. The parameter $G_{\mathrm{nom}}$ represents the nominal gain normalization factor, which is fixed at $10$ (via the \texttt{NOM\_GAIN} keyword) to conveniently yield an energy scaling of $10\mathrm{~eV}$ per unbinned PI channel. The physical coefficients $G_{C0}$ through $G_{C5}$ dictate the slope and intercept of the E-C relation.

In practice, extensive on-ground calibrations revealed that the WXT CMOS sensors exhibit remarkable spatial uniformity \citep[e.g.,][]{Cheng2024, 2025ChengWXTcalib}. The variance in the gain coefficients across the entire detector plane is constrained to a negligible $1-2\%$. Consequently, the spatial gradient parameters ($G_{C1}, G_{C2}, G_{C4}, G_{C5}$) are initially set to zero in the operational database. The transformation thus simplifies to a uniform linear scaling across the focal plane, dominated entirely by the baseline gain ($G_{C0}$) and the intercept offset ($G_{C3}$),
\begin{equation}
    \mathrm{PI} = \frac{ \mathrm{PHA} \cdot G_{C0} + G_{C3}}{10}
\end{equation}

During the in-orbit operation, however, the effective gain is highly susceptible to temporal degradation from cumulative radiation damage and long-term variations in the temperature of the detector module. To robustly track this continuous physical evolution, the gain files are structured as a multi-dimensional interpolation grid rather than a static look-up table, defined by two primary axes: \texttt{TIME} and \texttt{CMOSTEMP}. Specifically, the \texttt{TIME} axis serves as the direct proxy for capturing cumulative space radiation damage and long-term aging. In the data analysis pipeline, when an updated calibration result is appended as a new row in the FITS table at timestamp \texttt{TIME}$_i$, subsequent scientific observations automatically adopt these parameters. Crucially, the latest in-orbit calibration campaign conducted from late 2025 to early 2026 revealed no statistically significant changes in either the detector gain or the spectral energy resolution, demonstrating that the accumulated radiation damage to the WXT CMOS array has been practically negligible to date \citep{cheng2026epwxt}.

Regarding the thermal axis, the current baseline release (Version 1.0) pre-allocates three reference temperature nodes to strictly comply with the HEASARC interpolation standard framework; currently, all three nodes are populated with identical coefficients baselined at $-30^\circ\mathrm{C}$ (i.e., [$-30^\circ\mathrm{C}$, $-30^\circ\mathrm{C}$, $-30^\circ\mathrm{C}$]). This configuration is supported by the in-orbit operations and on-ground experiments. First, telemetry monitoring since the launch in January 2024 shows that the massive CMOS arrays have operated stably at their standard operating temperature of $-30^\circ\mathrm{C}$ without notable thermal elevation. Second, on-ground characterizations demonstrated that the CMOS gain depends weakly on thermal drifts. 
Specifically, comprehensive tests conducted across the entire array of 48 CMOS sensors comparing the nominal $-30^\circ\mathrm{C}$ baseline against $+20^\circ\mathrm{C}$ (characterized directly under ambient room temperature conditions without active cryogenic cooling) revealed a systematic mean relative gain shift of merely $(-1.8 \pm 0.2)\%$, indicating that the gain perturbation is quite slight even across a severe $50^\circ\mathrm{C}$ thermal gradient.
Should future telemetry reveal permanent thermal deviations, updated coefficients for different temperature nodes (e.g., $0^\circ\mathrm{C}$ or $+20^\circ\mathrm{C}$) can be directly ingested into these pre-allocated slots without altering the pipeline software architecture.

In summary, the WXT data analysis software determines the optimal E-C relation for a specific photon event by performing a linear interpolation. This algorithm continuously interpolates across the closest tabulated temporal epochs and thermal nodes, thereby ensuring resilient and precise energy calibration regardless of the satellite's in-orbit thermal cycling or long-term radiation exposure.

\subsubsection{Grade Calibration File}
\label{sec:grade}

When incident X-ray photons interact with the CMOS sensors, the resulting charge clouds frequently diffuse across multiple adjacent pixels. To systematically classify these charge-splitting morphologies, the CALDB incorporates specialized grade definition files, named as \texttt{epwgradeYYYYMMDDvNNN.fits}. Structurally, these files comprise a standard null primary header followed by a \textbf{\texttt{GRADES}} binary table extension. This table comprehensively maps various event topologies utilizing two fundamental columns: \texttt{GRADEID} (an integer index ranging from 0 to 32) and \texttt{GRADE}. The \texttt{GRADE} column encodes the specific topological pattern within a $3 \times 3$ pixel matrix centered on the local charge maximum. The structural configuration of this matrix is defined by a ternary logic: \texttt{1} designates an adjacent pixel exceeding the split threshold, \texttt{2} indicates a pixel falling below this threshold, and \texttt{0} acts as a `don't care' wildcard, accommodating flexible pattern matching.

For EP-WXT, the matrix framework is identical to the Photon Counting (PC) mode definitions historically employed by the \textit{Swift}/XRT mission\footnote{\url{https://swift.gsfc.nasa.gov/analysis/xrt_swguide_v1_2.pdf}}.
In practice, standard scientific analyses predominantly rely on the canonical subset spanning \texttt{GRADEID} 0 through 12. Within this convention, grade 0 isolates pure single-pixel events, grades 1--4 represent double splits, grades 5--8 correspond to triple splits, and grades 9--12 denote quadruplet charge distributions.

\subsubsection{Mirror Effective Area Calibration File}
\label{sec:effarea}

The intrinsic photon-collecting capability of the WXT optics is quantified by the mirror effective area files, conventionally named as \texttt{epw<cmos\_number>\_effareaYYYYMMDDvNNN.fits}. These files specifically profile the on-axis effective collecting area of the telescope as a direct function of incident energy. 
We note that the variation of the effective area at off-axis positions is independently parameterized within the vignetting calibration files.

The effective area baseline is rigorously derived from comprehensive ray-tracing simulations. This utilizes a customized framework meticulously tailored to the specific structural geometry and coating properties of the EP-WXT payload (Zhang et al., in preparation). To empower the data analysis software to accurately interpolate the mirror's theoretical transmission efficiency for any detected photon across the operational bandpass, the energy domain is evaluated with a fine resolution step of $5\mathrm{~eV}$. Structurally, the FITS file incorporates a standard null primary header followed by a binary table extension (named \textbf{\texttt{EFFAREA}}) constructed with the following two columns:
\begin{itemize}
    \item \texttt{ENERGY}: The incident photon energy scale, uniformly sampled in $5\mathrm{~eV}$ increments (expressed in eV);
    \item \texttt{EA}: The corresponding on-axis effective area of the mirrors (expressed in $\mathrm{cm}^2$).
\end{itemize}

\subsubsection{Filter Transmission Calibration File}
\label{sec:transmission}

\begin{figure*}[!htbp]
    \centering
    \includegraphics[width=0.49\textwidth]{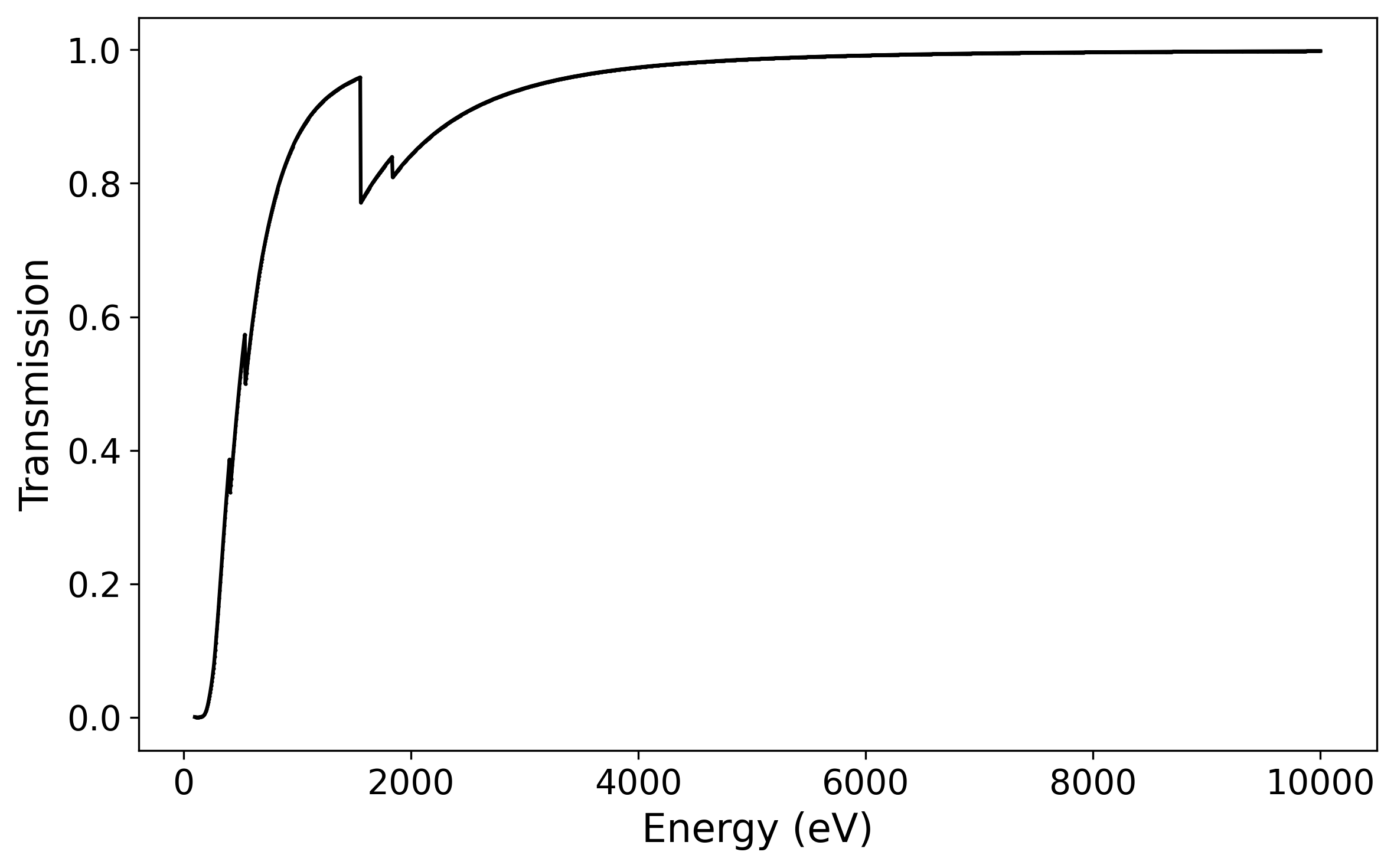}
    \includegraphics[width=0.49\textwidth]{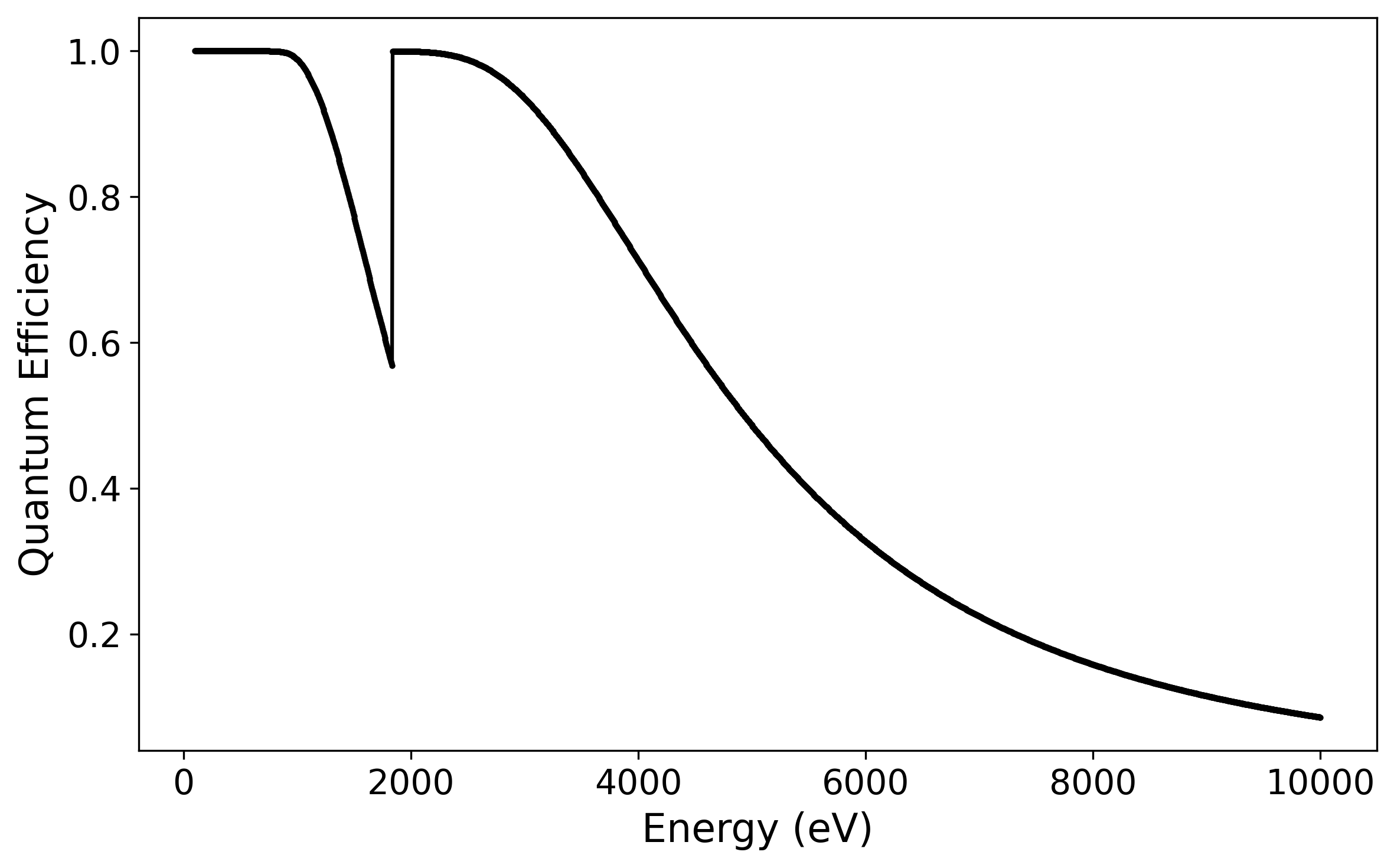}
    \caption{(\textit{Left} Panel) The filter transmission as a function of photon energy adopted for CMOS 1. (\textit{Right} Panel) The QE as a function of photon energy adopted for all CMOS sensors.}
    \label{fig:ftrans_and_qe}
\end{figure*}

The CMOS sensor on board the WXT payload is coated by a thin Aluminium filter to block optical light \citep[e.g.,][]{Wu2022, Wu2023pasp2}. 
The values of the filter transmission are derived from ray-tracing simulations and verified through on-ground calibration experiments. 
We maintain the transmission data in this calibration file, with the name of \texttt{epw<cmos\_number>\_ftransYYYYMMDDvNNN.fits}. 
The file format consists of an empty primary header and a binary table extension (named \textbf{\texttt{TRANSMISSION}}) with two columns, \texttt{ENERGY} and \texttt{TRANSMIS}. The first contains the energy values where the filter transmission was evaluated and the second contains the corresponding transmission value. The energy column ranges between 0.1--10 keV, with a step of $5\mathrm{~eV}$. The transmission curve for CMOS 1 is presented in the \textit{Left} panel of Figure \ref{fig:ftrans_and_qe}, as an example.

\subsubsection{Quantum Efficiency File}
\label{sec:qe}

The quantum efficiency files contain the quantum efficiency (QE) of the CMOS as a function of energy and different grades.
The name of the file is \texttt{epw<cmos\_number>\_qeYYYYMMDDvNNN.fits}.
The file format consists of a standard null primary header and a binary table extension (named \textbf{\texttt{QE}}). All files contain a column named \texttt{ENERGY} and a variable number of columns containing the quantum efficiency for different grade selections. The quantum efficiency is reported for grades 0, 0--4, and 0--12.
The values of the QE are generated by simulations and verified through on-ground calibrations. The QE curve for all CMOS detectors is presented in the \textit{Right} panel of Figure \ref{fig:ftrans_and_qe}.

\subsection{Calibration Product Files}
\label{sec:cpf}

The Calibration Product Files (CPF) represent high-level, analysis-ready products generated by synthesizing various fundamental instrumental and electronic parameters derived from the BCFs. While BCFs store the native characteristics of the instrument hardware, CPFs directly parameterize the overall macroscopic optical and detector responses such as energy redistribution and on-axis effective area response. These files are specifically formatted to be directly ingested by standard high-level scientific analysis software to transform observed event lists into physically calibrated spectra, light curves, and images.

\subsubsection{Response Matrix Function}
\label{sec:rmf}

The Response Matrix Function (RMF) defines the probability that an incident X-ray photon with a given energy will be registered in a specific detector channel. Combined with the ARF (Section \ref{sec:arf}), the RMF provides the necessary transformation between the true astrophysical spectrum and the detected instrumental signal. Beyond a simple energy-to-channel mapping, this matrix accounts for the detector's complex physical responses. It models the energy resolution, as well as distinct spectral features like low-energy shoulders and fluorescent escape peaks, which arise from partial charge collection and photon escape, respectively. 
These matrices are computed for the specific event grades and grade ranges considered `good' for the data analysis, and applied to spectra binned in PI channels.
Commonly adopted grade selections in scientific analysis include grade=\texttt{0}, \texttt{0--4}, and \texttt{0--12}. The corresponding file names for single grade selection is \texttt{epw<cmos\_number>\_<datamode><grade>\_YYYYMMDDvNNN.rmf} where \texttt{<grade>} is the grade of validity, and \texttt{epw<cmos\_number>\_<gradelow>to<gradehigh>\_YYYYMMDDvNNN.rmf} for grade range selections where \texttt{<gradelow>} is the lower grade of validity and \texttt{<gradehigh>} is the higher grade of validity.

Structurally, the WXT RMF file contains a standard null primary header followed by two binary table extensions: \textbf{\texttt{MATRIX}} and \textbf{\texttt{EBOUNDS}}. To efficiently store the sparse response array, the \textbf{\texttt{MATRIX}} extension utilizes the following standard columns:
\begin{itemize}
    \item \texttt{ENERG\_LO}: The lower boundary of the incident energy bin;
    \item \texttt{ENERG\_HI}: The upper boundary of the incident energy bin;
    \item \texttt{N\_GRP}: The number of continuous channel subsets for the given energy;
    \item \texttt{F\_CHAN}: The starting channel index for each respective subset;
    \item \texttt{N\_CHAN}: The total number of channels contained within each subset;
    \item \texttt{MATRIX}: The actual response probabilities for the specified channels.
\end{itemize}
The subsequent \textbf{\texttt{EBOUNDS}} extension defines the nominal energy scale of the detector. It comprises three columns: \texttt{CHANNEL} (PI channels, ranging from 1 to 1024), \texttt{E\_MIN}, and \texttt{E\_MAX}, which represent the physical lower and upper limits (in keV) assigned to each channel.

\begin{figure*}[!htbp]
    \centering
    \includegraphics[width=\textwidth]{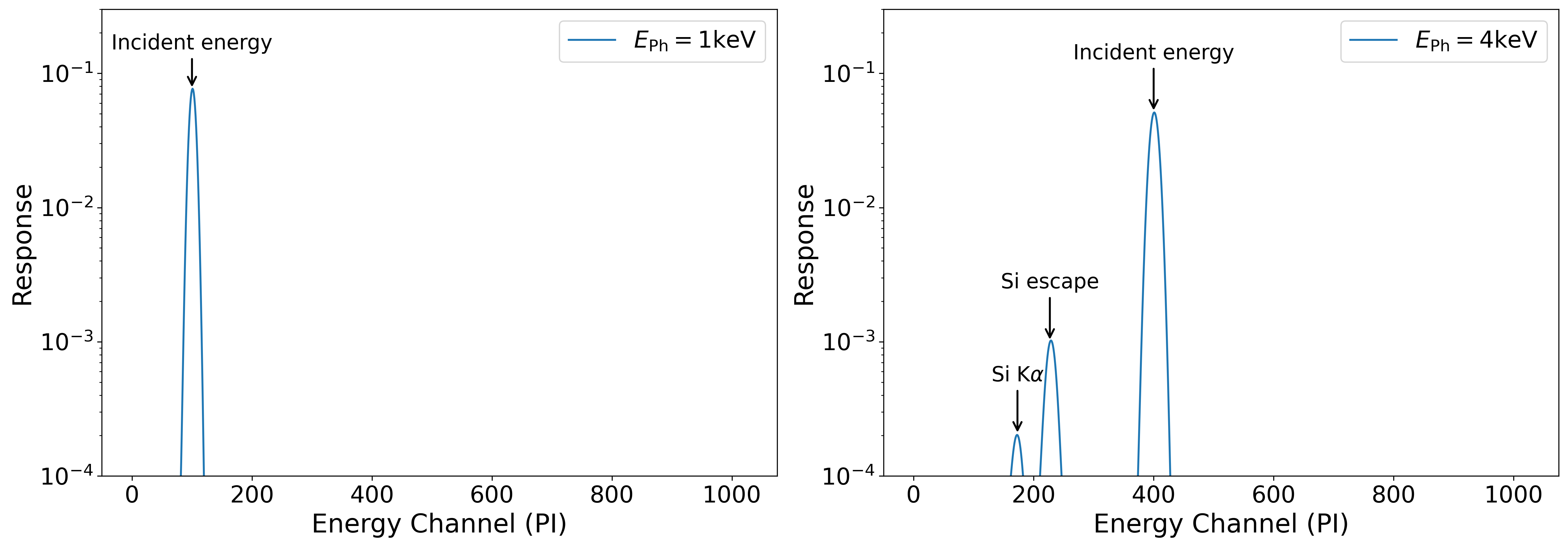}
    \caption{The CMOS response array for an incident photon with energy ranging from $1\mathrm{~keV}$ to $1.005\mathrm{~keV}$ (\textit{Left} Panel) and $4\mathrm{~keV}$ to $4.005\mathrm{~keV}$ (\textit{Right} Panel). In both scenarios, the channel corresponding to the maximum peak aligns with the primary incident energy. For the higher-energy $4\mathrm{~keV}$ photon, the secondary structures corresponding to the Silicon escape peak and the Si K$\alpha$ fluorescence line are explicitly annotated. Grade \texttt{0--12} selection is adopted.}
    \label{fig:rmf}
\end{figure*}

As an example, Figure \ref{fig:rmf} displays the extracted response profiles from the CALDB for incident photons at $1\mathrm{~keV}$ and $4\mathrm{~keV}$. Several peaks in the response distribution are denoted with arrows, including the maximum peak at the incident photon energy ($E_{\mathrm{incident}}$), and two less prominent ones at the energies of the Si K$\alpha$ line ($1.74\mathrm{~keV}$) and Si escape peak ($E = E_{\mathrm{incident}} - 1.74$) in the latter case.

In orbit, the operational performance of the RMF is routinely verified using observations of the line-rich supernova remnant Cassiopeia A (Cas A). Spectral monitoring accumulated over approximately 2 years of operations demonstrates that the energy resolution has remained highly stable, exhibiting no statistically measurable broadening \citep{cheng2026epwxt}. Consequently, Version 1.0 of the CALDB retains the static redistribution matrices derived from the on-ground calibration campaigns. From a lifecycle perspective, the RMF update mechanism is governed by an epoch-based trigger: a new version will only be generated and ingested when celestial line monitoring reveals statistically significant spectral degradation.

\subsubsection{Ancillary Response File}
\label{sec:arf}

The Ancillary Response File (ARF) stored in the CALDB is the on-axis ARF for a standard extraction radius. Specifically, it defines the on-axis effective area as a function of energy, incorporating the combined physical effects of the MPO mirror effective area, the optical filter transmission, and the CMOS detector QE.
The ARF generator tool is distributed within the data analysis software, therefore users can create the ARF accordingly with the extracted spectrum. 

The file name is \texttt{epw<cmos\_number>\_<gradelow>to<gradehigh>\_YYYYMMDDvNNN.arf}, where \texttt{<gradelow>} is the lower grade of validity and \texttt{<gradehigh>} is the higher grade of validity. Three types of grade selection are employed, including \texttt{0}, \texttt{0--4} and \texttt{0--12}, in accordance with the grade selection of the RMF. The file format consists of a standard null primary header and a binary table extension named \textbf{\texttt{SPECRESP}}. The table extension contains three standard columns:
\begin{itemize}
    \item \texttt{ENERG\_LO}: The lower boundary of the incident energy bin;
    \item \texttt{ENERG\_HI}: The upper boundary of the incident energy bin;
    \item \texttt{SPECRESP}: The on-axis quantum efficiency and filter transmission corrected effective area for the specified channels. The unit is $\mathrm{cm}^2$.
\end{itemize}
The energy range is $0.1$--$10\mathrm{~keV}$, with a step of $5\mathrm{~eV}$. 

Regarding the in-orbit verification and lifecycle management of the ARF, the effective area has been calibrated through successive campaigns using the canonical X-ray standard candle, the Crab Nebula \citep{cheng2026epwxt}. Specifically, spectral fits to Crab observations yield power-law indices, column densities and flux normalizations in good agreement with established canonical values, verifying the baseline effective area post-launch. Our latest calibration observations ($\sim2$~years post-launch) indicate that a minority of detector modules have begun to exhibit a slight effective area deterioration in the low-energy band. This gradual trend is a well-documented space engineering phenomenon attributable to time-dependent molecular contamination buildup onto cold optical surfaces, identical to the long-term behaviors observed in other space missions such as \textit{Chandra} and \textit{XMM-Newton}. It should be noted that Version 1.0 represents the clean early-mission baseline and explicitly excludes this slow decay. A dedicated calibration campaign scheduled for late 2026 will further validate and parameterize this low-energy attenuation, which will be pushed as an epoch-updated ARF in the subsequent major release (CALDB Version 2.0).

\subsubsection{Vignetting Calibration File}
\label{sec:vignetting}

\begin{figure*}[!htbp]
    \centering
    \includegraphics[width=\textwidth]{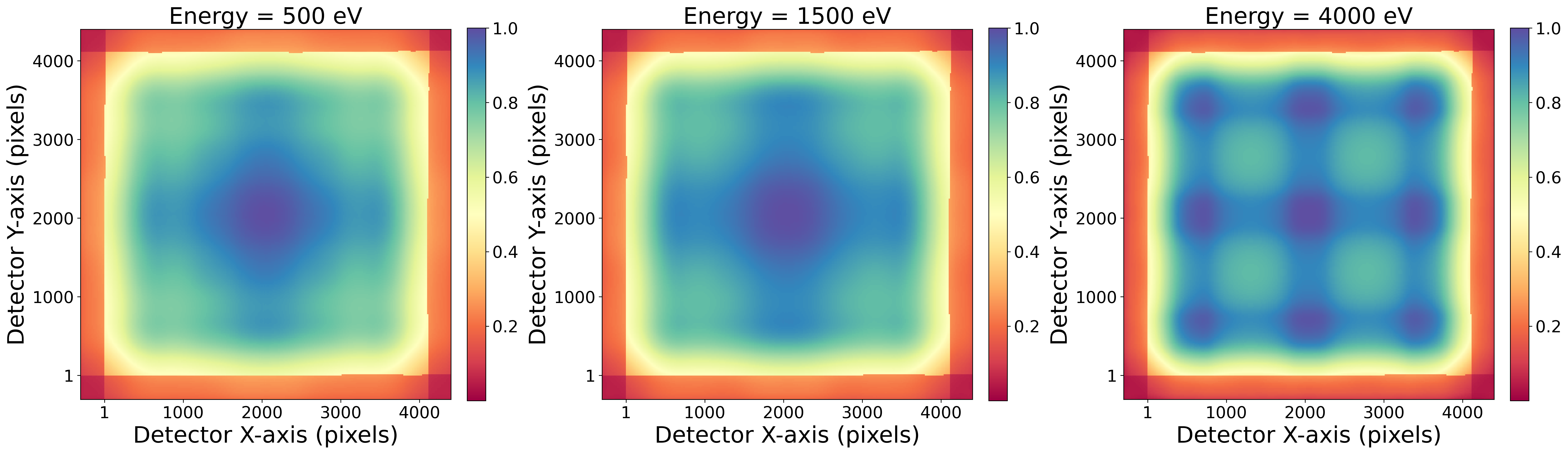}
    \caption{The vignetting correction value mapping for a CMOS detector at different energies of incident photons (\textit{Left} Panel: $0.5\mathrm{~keV}$, \textit{Middle} Panel: $1.5\mathrm{~keV}$, \textit{Right} Panel: $4\mathrm{~keV}$).}
    \label{fig:vignetting}
\end{figure*}

The WXT vignetting calibration file stores the ratio between the effective area at a given position of a CMOS and that at the detector center. This ratio also depends on energy. The filename is \texttt{epw<cmos\_number>\_vignYYYYMMDDvNNN.fits}.
This calibration file consists of a standard null primary header and several \textbf{\texttt{IMAGE}}-type extensions. These extensions contain the value of the vignetting factor at the energy indicated by the suffix of the extension name (e.g., extension \textbf{\texttt{VIG\_200}} corresponds to the vignetting data of $200\mathrm{~eV}$), as well as the keyword \texttt{CBD10001} in each extension. 
In each \textbf{\texttt{IMAGE}} extension, the original pixel of a CMOS is binned by a factor of 16 to record the positional dependence of the vignetting factor in a $295 \times 295$ array (we note that the practically mapping coordinates are in range of [-304, 4400] for both axes).
The data analysis software maps the position of the source to the corresponding pixel, and then the value of the vignetting factor at a given energy is interpolated by the values stored in the extensions.
An example of the vignetting correction map is presented in Figure \ref{fig:vignetting}. The three panels correspond to photons from low to high energies. The vignetting map is generated by ray-tracing simulations and verified by on-ground calibration experiments.

\subsubsection{PSF Correction File}
\label{sec:psf}

The profile of the WXT PSF depends on the photon energy as well as the position of the focal spot locating in the FoV \citep[see e.g.,][]{2017ZhaoDH_simulation, 2022ZhangChenApJL, Cheng2024, 2025ChengWXTcalib}. 
The PSF correction factor is defined as the ratio between the flux in a circular aperture with a radius of $9.2^{\prime}$
\footnote{This is a standardized extraction shape for the source spectrum and light curve. In practice, under this aperture choice the modeled effective area matches with the on-ground calibration result across the WXT effective energy band.} centered at the focus spot and that of the entire PSF within the FoV of a CMOS (note that the cruciform arms may fall out of the FoV). 

\begin{figure*}[!htbp]
    \centering
    \includegraphics[width=\textwidth]{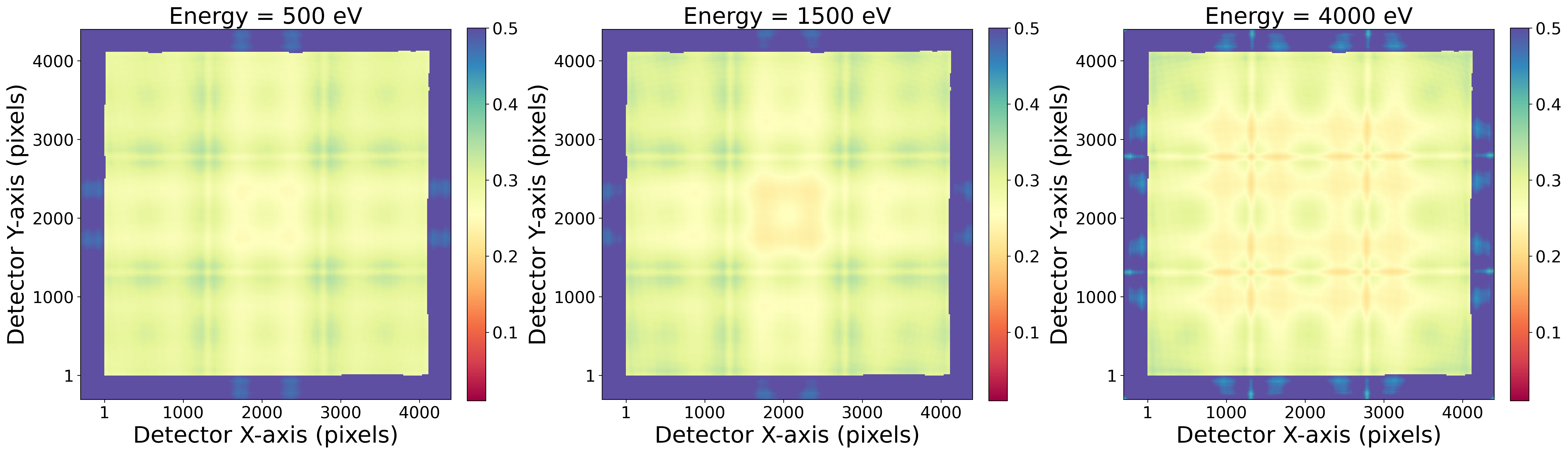}
    \caption{Mapping of the PSF correction coefficient for a CMOS detector at different energies of incident photons (\textit{Left} Panel: $0.5\mathrm{~keV}$, \textit{Middle} Panel: $1.5\mathrm{~keV}$, \textit{Right} Panel: $4\mathrm{~keV}$).}
    \label{fig:psf}
\end{figure*}

The PSF correction file is named as \texttt{epw<cmos\_number>\_psfYYYYMMDDvNNN.fits}. This CALDB file consists of a null standard primary HDU and several \textbf{\texttt{IMAGE}} extensions. These extensions contain the value of the PSF correction factor at the energy indicated by the suffix of the extension name (e.g., extension \textbf{\texttt{PSF\_200}} corresponds to the correction data of $200\mathrm{~eV}$), as well as the keyword \texttt{CBD10001} in each extension. 
In each \textbf{\texttt{IMAGE}} extension, the original pixel of a CMOS detector is binned by a factor of 16 to record the positional dependence of the PSF correction factor in a $295 \times 295$ array (we note that the practically mapping coordinates are in range of [-304, 4400] for both axes). 
An example of the PSF correction matrix at three different energies ($500\mathrm{~eV}$, $1500\mathrm{~eV}$ and $4000\mathrm{~eV}$) for one single CMOS detector is presented in Figure \ref{fig:psf}. In general, the CMOS plane is divided into $3 \times 3$ separate regions, each corresponding to one of the anterior 9 MPO chips. 
This correction map is generated by ray-tracing simulations.
The data analysis software maps the position of the source to the corresponding pixel, then the value of the PSF correction factor at a given energy is interpolated by the values stored in the extensions.

\subsubsection{Background Spectra}
\label{sec:background}

The background spectra maintained in CALDB are used to subtract the background in the spectra when the extraction region of the background is not available. The filename is \texttt{epw<cmos\_number>\_bkgYYYYMMDDvNNN.pha}.
The file consists of a standard null primary header and a binary table extension named \textbf{\texttt{SPECTRUM}}. The table extension contains four standard columns:
\begin{itemize}
    \item \texttt{CHANNEL}: The PI channel ranging from 1--1024;
    \item \texttt{COUNTS}: The estimated background counts within the channel;
    \item \texttt{QUALITY}: Data quality. All rows are set to 0.
    \item \texttt{GROUPING}: Information of the data grouping. All rows are set to zero.
\end{itemize}

The background spectrum is initially generated based on Monte Carlo simulations, utilizing the methods detailed in \citet{2018ZDH_WXTbackground}. 
With the operation of the mission, the spectrum will be updated using the WXT all-sky map data (Pan et al., in preparation).

\section{Summary}
\label{sec:summary}

In this paper, we present the design and implementation of the calibration database for the WXT payload on board the EP satellite, which serves as a fundamental prerequisite for enabling precise and reliable scientific data analysis. In particular, WXT's innovative lobster-eye micro-pore optics (MPO) and its large-format array of 48 CMOS sensors present unique calibration challenges, which necessitated a tailored approach to the design and architectural organization of the CALDB. 
A comprehensive, HEASARC-compliant CALDB has been developed for the payload, systematically organizing basic calibration files and high-level products.
Following the successful launch and the commencement of scientific operations, the EP science team has effectively utilized the WXT CALDB in conjunction with the data analysis pipeline. 
The wealth of mission outputs across diverse astronomical fields, characterized by high scientific impact and citation rates \citep{Lewis2026arxiv}, has clearly demonstrated the robustness and reliability of the calibration database in managing and interpreting the complex data streams generated by the unprecedented large FoV of the 12 independent WXT modules.

The structures and contents described in this paper reflect the WXT CALDB Version 1.0, providing the foundational calibration basis for the public release of EP data scheduled for late-2026. To benchmark its precision, Table ~\ref{tab:calib_summary} summarizes the primary in-orbit calibration results obtained during the initial two and a half years of operations \citep{cheng2026epwxt}.
Specifically, the in-orbit PSF demonstrates a focal spot full-width-at-half-maximum (FWHM) ranging from $3^{\prime}$ to $6^{\prime}$ across $\sim90\%$ of the FoV (with a median of $\sim4.3^{\prime}$). The source positioning accuracy reaches $1.3^{\prime}$ at the $90\%$ confidence level, following the refinement of celestial-to-detector coordinate transformation matrices and the modeling of residual non-linear spatial distortions.
The in-orbit effective area agrees with model predictions and on-ground measurements, exhibiting a systematic uncertainty of $\le10\%$ ($90\%$ C.L.) in the $0.5$--$4\mathrm{~keV}$ soft X-ray band. Nevertheless, a gradual low-energy attenuation ($0.4$--$0.6\mathrm{~keV}$) is observed, which exhibits a diverse module-to-module behavior and reaches $\sim30\%$--$40\%$ in a small subset of detectors. 
Meanwhile, the energy scale and spectral resolution of the CMOS detectors remain largely stable, with the typical in-orbit spectral resolution FWHM concentrated within $\sim130$--$150\mathrm{~eV}$ in the $\sim2\mathrm{~keV}$ energy regime, which is in line with expectations interpolated from on-ground calibrations.

\begin{table}[htbp]
  \centering
  \renewcommand{\arraystretch}{1.2} 
  \caption{Summary of primary WXT in-orbit baseline calibration precision (CALDB Version 1.0).}
    \begin{tabular}{@{} >{\raggedright\arraybackslash}p{3.5cm} >{\raggedright\arraybackslash}p{5.8cm} >{\raggedright\arraybackslash}p{4.2cm} @{}}
    \toprule
    \textbf{Calibration Item} & \textbf{Baseline Performance (V1.0)} & \textbf{Verification Source} \\
    \midrule
    \textbf{Gain} (\texttt{GAIN}) & Highly stable post-launch; line centroid offsets $\le$ a few PHA channels. & Cassiopeia A (Cas A) \\
    \addlinespace
    \textbf{Energy Resolution} (\texttt{RMF}) & FWHM $\sim130$--$150\mathrm{~eV}$ in the $\sim2\mathrm{~keV}$ energy regime. & Cas A \\
    \addlinespace
    \textbf{Spatial Resolution} (\texttt{PSF}) & Median FWHM $\sim4.3^{\prime}$ (typically ranges from $3^{\prime}$ to $6^{\prime}$ across $\sim90\%$ of FoV). & Crab Nebula / Scorpius X-1 \\
    \addlinespace
    \textbf{Source Localization} (\texttt{TELDEF}) & Systematic error radius $R_{90}=1.3^{\prime}$ ($90\%$ confidence level). & Source Catalogs (NED positions as reference) \\
    \addlinespace
    \textbf{Effective Area} (\texttt{ARF}) & Systematic uncertainty $\le10\%$ ($90\%$ C.L.) in the $0.5$--$4\mathrm{~keV}$ band. & Crab Nebula \\
    \bottomrule
    \end{tabular}
  \label{tab:calib_summary}
\end{table}

Given the novel spaceborne employment of MPO and CMOS technologies, tracking the long-term evolution of instrumental performance is critical for sustained, high-impact scientific output. Moving forward, our database maintenance will be governed by a multi-tiered operational protocol designed to ensure continuous accuracy while preserving pipeline stability. Specifically, for basic detector properties such as bias maps and bad pixels, automated tracking is implemented via the routine acquisition and analysis of dark frames. Regarding the PSF (\texttt{PSF}), effective area (\texttt{ARF}) and detector energy response (specifically, \texttt{GAIN} and \texttt{RMF}), dedicated calibration observations of canonical standard candles (e.g., Cas A and the Crab Nebula) will be scheduled on an annual basis to monitor potential radiation damage or contamination buildup; concurrently, we maximize the reuse of routine sky survey data to enrich the calibration sample. For astrometric calibration, rather than relying on dedicated calibration pointings—which would be highly time-consuming to map the PSF focal spot positions across different incident angles—we will continuously track the evolution of localization precision and opportunistically refine the coordinate transformation matrices using routine scientific observations of source-dense fields, such as the Galactic center, bulge, and plane, thereby significantly enhancing overall calibration efficiency. Finally, we note that Version 1.0 represents the clean early-mission baseline, with no inclusion of the slow localized low-energy effective area deterioration. This will be parameterized in the subsequent major release (CALDB Version 2.0).

In the future, we plan to expand the scope of the current calibration parameters, develop and optimize automated calibration-derivation tools, and explore advanced operational functionalities. These future endeavors, including real-time telemetry-driven data updates and intelligent spatial background modeling support, will further enhance the practicality and efficiency of the WXT CALDB, ensuring optimal precision for the discovery and characterization of cosmic X-ray transients.

\begin{acknowledgements}
We thank the anonymous referee for very valuable comments that help improve the paper. This work is based on the data obtained with the Einstein Probe (EP), also known as `Tianguan'. EP is led by the Chinese Academy of Sciences, in collaboration with the European Space Agency, the Max Planck Institute for Extraterrestrial Physics (Germany), and the Centre National d'Études Spatiales (France).
This work is supported by National Key R\&D Program of China No. 2025YFF0511100.
This work is supported by the National Natural Science Foundation of China (Grant Nos. 12333004, 12433005, 12473100, 12273073), and the Strategic Priority Research Program of the Chinese Academy of Sciences (Grant No.XDB0550200).
\end{acknowledgements}

\bibliographystyle{raa}
\bibliography{bibtex}

\end{document}